\documentclass{emulateapj}

\newcommand{\eg}{e.\,g.\,}

\newcommand{\SiII}{\ion{Si}{2}}
\newcommand{\SII}{\ion{S}{2}}

\newcommand{\iprime}{i\arcmin}
\shorttitle{SNLS Gemini Spectra}
\shortauthors{Howell}

\begin{document}

\title{Gemini Spectroscopy of Supernovae from SNLS: Improving High 
Redshift SN Selection and Classification
}
\author{D.~A.~Howell\altaffilmark{1},
M.~Sullivan\altaffilmark{1},
K.~Perrett\altaffilmark{1}, 
T.~J.~Bronder\altaffilmark{2}, 
I.~M.~Hook\altaffilmark{2}, 
P.~Astier\altaffilmark{3}, 
E.~Aubourg\altaffilmark{4,5},
D.~Balam\altaffilmark{6}, 
S.~Basa\altaffilmark{7},
R.~G.~Carlberg\altaffilmark{1},
S.~Fabbro\altaffilmark{8},
D.~Fouchez\altaffilmark{9}, 
J.~Guy\altaffilmark{3}, 
H.~Lafoux\altaffilmark{5},
J.~D.~Neill\altaffilmark{6}, 
R.~Pain\altaffilmark{3}, 
N.~Palanque-Delabrouille\altaffilmark{5},
C.~J.~Pritchet\altaffilmark{6},
N.~Regnault\altaffilmark{3}, 
J.~Rich\altaffilmark{5}, 
R.~Taillet\altaffilmark{3},
R.~Knop\altaffilmark{10},
R.~G.~McMahon\altaffilmark{11},
S.~Perlmutter\altaffilmark{12},
N.~A.~Walton\altaffilmark{11}
}
                                                          
\altaffiltext{1}{Department of Astronomy and Astrophysics, University of
Toronto, 60 St. George Street, Toronto, ON M5S 3H8, Canada}
\altaffiltext{2}{University of Oxford Astrophysics, Denys Wilkinson
  Building, Keble Road, Oxford OX1 3RH, UK}
\altaffiltext{3}{LPNHE, CNRS-IN2P3 and University of
Paris VI \& VII, 75005 Paris, France}
\altaffiltext{4}{APC, 11 Place Marcelin Berthelot, 75231 Paris Cedex 05, France}
\altaffiltext{5}{DSM/DAPNIA, CEA/Saclay,
91191 Gif-sur-Yvette Cedex, France}
\altaffiltext{6}{Department of Physics and Astronomy, University of
Victoria, PO Box 3055, Victoria, BC V8W 3P6, Canada}
\altaffiltext{7}{LAM, BP8, Traverse du Siphon, 13376 Marseille Cedex 12, France}
\altaffiltext{8}{CENTRA - Centro Multidisciplinar de Astrof\'{\i}sica, IST, Avenida Rovisco
Pais, 1049 Lisbon, Portugal}
\altaffiltext{9}{CPPM, CNRS-Luminy, Case 907, 13288 Marseille Cedex 9, France}
\altaffiltext{10}{Department of Physics and Astronomy, Vanderbilt
  University, VU Station B 351807, Nashville, TN 37235-1807 USA}
\altaffiltext{11}{Institute of Astronomy, University of Cambridge,
  Madingley Road, Cambridge CB3 0HA, UK}
\altaffiltext{12}{University of California, Berkeley and
Lawrence Berkeley National Laboratory, Mail Stop 50-232, Lawrence Berkeley National Laboratory, 1
Cyclotron Road, Berkeley CA 94720 USA}

\begin{abstract}  
  We present new techiques for improving the efficiency of supernova
  (SN) classification at high redshift using 64 candidates observed at
  Gemini North and South during the first year of the Supernova Legacy
  Survey (SNLS).  The SNLS is an ongoing five year project with the
  goal of measuring the equation of state of Dark Energy by
  discovering and following over 700 high-redshift SNe Ia using data
  from the Canada-France-Hawaii Telescope Legacy Survey.  We achieve
  an improvement in the SN Ia spectroscopic confirmation rate: at
  Gemini 71\% of candidates are now confirmed as SNe Ia, compared to
  54\% using the methods of previous surveys.  This is despite the
  comparatively high redshift of this sample, where the median SN Ia
  redshift is $z=0.81$ ($0.155 \leq z \leq 1.01$).  These improvements
  were realized because we use the unprecedented color coverage and
  lightcurve sampling of the SNLS to predict whether a candidate is an
  SN Ia and estimate its redshift, before obtaining a spectrum, using
  a new technique called the ``SN photo-z.''  In addition, we have
  improved techniques for galaxy subtraction and SN template $\chi^2$
  fitting, allowing us to identify candidates even when they are only
  15\% as bright as the host galaxy.  The largest impediment to SN
  identification is found to be host galaxy contamination of the
  spectrum -- when the SN was at least as bright as the underlying
  host galaxy the target was identified more than 90\% of the time.
  However, even SNe on bright host galaxies can be easily identified
  in good seeing conditions.  When the image quality was better than
  0.55\arcsec , the candidate was identified 88\% of the time.  Over
  the five-year course of the survey, using the selection
  techniques presented here we will be able to add $\sim 170$ more
  confirmed SNe Ia than would be possible using previous methods.

\end{abstract}


\section{Introduction}
Type Ia supernovae (SNe Ia) have been used as standardized candles to
discover the acceleration of the universe 
\citep{1998AJ....116.1009R,1999ApJ...517..565P}.  Programs are now 
underway to characterize the Dark Energy driving this
expansion by measuring its average equation of state, 
$\langle w \rangle = p/\rho$.  ESSENCE
aims to discover and follow-up $\approx$ 200 SNe Ia for 
this purpose \citep{2005AJ....129.2352M},
while the goal of the SNLS (Supernova Legacy 
Survey) is to obtain 700 well observed SNe Ia in the redshift
range 0.2 to 0.9 \citep{2004astro.ph..6242P,2004astro.ph.10594S}. 

The SNLS uses the Canada-France-Hawaii Telescope Legacy Survey
(CFHTLS) imaging data for SN discoveries and lightcurves.  Over the
course of a year, four fields are imaged every four days (2--3 days
rest frame) during dark and gray time, in four filters (g\arcmin
r\arcmin i\arcmin z\arcmin ).  Since the same fields are continuously
imaged, this is a ``rolling search,'' where new SN candidates are
discovered throughout the month as lightcurves are being built.

The SNLS uses the Very Large Telescope (VLT), Keck and Gemini
telescopes to determine SNe types and redshifts.    The VLT has 
the shortest setup time, so it
handles the bulk of the lower redshift ($z<0.8$) candidates, as
well as some higher redshift targets.  Gemini-N and Gemini-S generally
observe the faintest, highest redshift targets ($z>0.6$), where the
nod-and-shuffle mode provides a reduction of sky line residuals in
the red part of the spectrum.  Finally, Keck provides essential
observations when the northernmost field cannot be observed by
the VLT.

Here we present the methods and spectroscopy from the Gemini
telescopes for the first year of operation of the SNLS.  Separate
science analyses on certain
spectroscopic features, as well as observations from other telescopes,
will be presented in upcoming papers.  

A note on our candidate names: the classical International 
Astronomical Union (IAU) naming convention for
supernovae does not work well for projects of the scale of the SNLS,
where there are more candidates than confirmed SNe.  Under the
traditional naming system, the candidates from high-redshift searches have an
internal (unofficial) name that they were given at the time of
discovery, then if they
are suspected SNe they are given one type of official IAU name, and if they are
confirmed SNe they are given a second type of official IAU name.  To avoid these
problems, we use a unified naming scheme for all SNLS discoveries,
regardless of whether they are confirmed to be SNe.
An example name is SNLS-04D3aa --- the first two digits are the year of
discovery, the second two are the field name (D1, D2, D3, D4), and the
final two are identifying letters.  We start with ``aa'' for the first
candidate discovered in a field in a given year.  The second is
``ab,'' then ``ac'' and so on, until the 27th candidate, which is
named ``ba.''  

\section{Candidate selection:  SN Photo-z}
\label{photozintro}

Historically, only $\sim 45-60\%$ of high-redshift SN candidates
observed spectroscopically in the range $0.2<z<1.0$ are confirmed as
Type Ia SNe \citep{2005A&A...430..843L,2005AJ....129.2352M}.  The rest
of the candidates are either identified or unidentified core collapse 
SNe or AGN, or are SNe Ia which have such high host contamination or 
low signal-to-noise (S/N) that they cannot be identified.  The 
exact percentage confirmed depends on
the design of the search and the redshift range studied (see section
\ref{photoz} for a detailed discussion).

\citet{2005A&A...430..843L} also found that the SN Ia 
confirmation rate was
higher for classical searches than it was for ``rolling'' searches.
A classical high-z SN search technique
\citep{1995ApJ...440L..41P} is optimized for the discovery of SNe Ia.
It involves taking a reference image after new moon to be subtracted
from a discovery image taken $\sim$3 weeks later (2 weeks restframe at
z=0.5).  Since the rise time of SNe~Ia is about 18 days, this
guarantees that the SNe discovered will be either rising or near
maximum light.  Cuts are made to retain only candidates that have
increased in brightness by an amount consistent with the rapid rise
of an SN~Ia.  

In ``rolling searches,'' such as the SNLS and ESSENCE, the field is
continuously imaged every few days in several filters.  SNe are
discovered very early, so there is no built-in bias toward the
discovery of SNe Ia (except that SNe Ia are usually brighter than core
collapse SNe at a given redshift).  

Rolling searches are typically limited not by the number of candidates
discovered, but by
the amount of spectoscopic confirmation time available, so techniques
must be developed to increase the percentage of targets confirmed as
SNe Ia by spectroscopy.  In principle, this should be easier in a
rolling search than in a classical search.  In the SNLS, SNe are almost
almost always discovered well before maximum light, so multicolor
information on the rising part of the lightcurve is available before
the candidates are observed spectroscopically.

Here we demonstrate the use of a technique to prioritize
candidate spectroscopy.  After each photometric data point is taken,
a candidate's lightcurve is fit under the assumption that it is an SN
Ia.  The lightcurve shape, phase, and redshift of the candidate are
allowed to vary in the fit, while the Milky Way reddening is fixed to
the values of \citet{1998ApJ...500..525S}.  Host reddening can be
solved for or held fixed, depending on the number of data points fit.
We do not use a fit until it has detections in at least two epochs 
in at least two filters, as well as at least one prior epoch of 
nondetections.  In practice, most candidates have detections on several 
epochs in at least three filters, and dozens of nondetections.
Significantly deviant objects are lowered in priority (for example,
Type II SNe and AGN are often too blue at early times
\citep[see][]{2005sullivan, 2004ApJ...600L.163R}), while rising 
candidates that are
consistent with being an SN Ia are given top priority for
spectroscopic observation.

Because the inferred color of the object depends on the redshift, one
product of the fitting process is an estimated redshift.  We call this
a ``supernova photometric redshift,'' or ``SN photo-z,'' since the
redshift is estimated from prior assumptions about the intrinsic
brightness of SNe Ia (0.2 mag of dispersion is allowed for in the
assumed SN Ia properties).  The SN photo-z is independent from galaxy
photometric redshifts, and is useful for determing the telescope and
instrument setup necessary to observe key SN Ia spectroscopic
features.  Note that galaxy photometric redshifts were not used in
assigning the follow-up priority for the targets considered in this paper.

\begin{figure}
\plotone{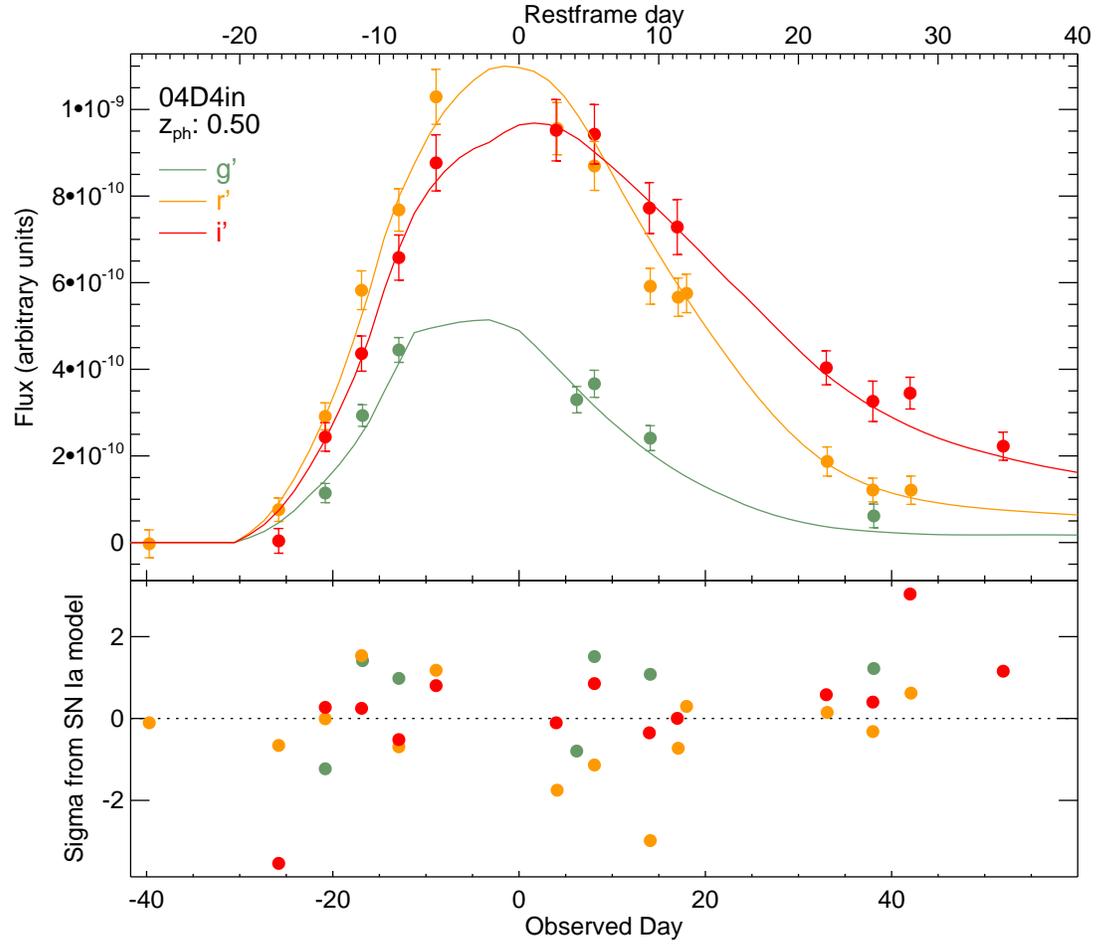}
\caption{SN Ia lightcurve fit to real-time data for an example SN candidate that was
  spectroscopically confirmed as an SN Ia.  Data are fit after each
  photometric point is acquired, solving for the redshift (the SN
  photo-z).  If the fit is consistent with a Ia the candidate is given
  high priority for spectroscopic follow-up.  The predicted redshift
  from the fit is $z=0.50$, the actual redshift from spectroscopy is
  0.516.\label{Iapz}}
\end{figure}

\begin{figure}
\plotone{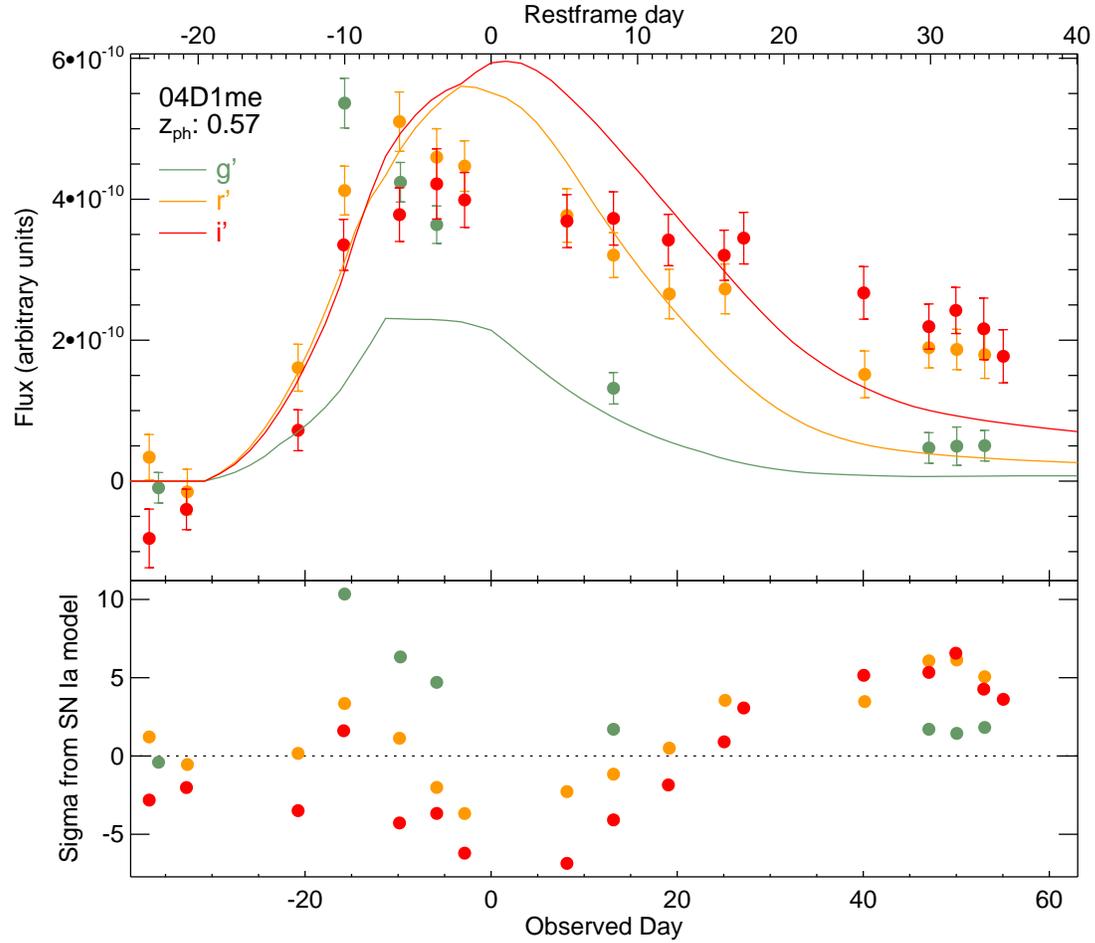}
\caption{A SN Ia lightcurve fit to an SN candidate that
  was eventually spectroscopically identified as a Type II-P.  Because
  the data are not a good fit to an SN Ia lightcurve, such candidates
  are lowered in priority for spectroscopic follow-up.  Even at early
  times it is apparent that the candidate is not an SN Ia because it
  is too blue.  The formal prediction for the SN photo-z (under the
  assumption that the candidate is an SN Ia) is 0.56.  The
  spectroscopic redshift is $z=0.256$ --- much lower because SNe II
  are $\sim 1.5$ mag fainter than SNe Ia at a given
  redshift.\label{IIpz}}
\end{figure}

Figures~\ref{Iapz} and \ref{IIpz} show examples of SN photo-z fits to
objects that were eventually confirmed to be Type Ia and Type II SNe,
respectively.  While we show the entire lightcurve fit here for
clarity, it is apparent even before maximum light that the candidate
in Figure~\ref{Iapz} is likely to be an SN Ia, while the candidate in
Figure~\ref{IIpz} is too blue at early times to be an SN Ia.  The
candidates are usually spectroscopically observed near maximum light,
so most SN photo-z fits are done using only the rising part of the
lightcurve.


Since we observe many of our highest redshift (and therefore
spectroscopically expensive) targets at Gemini, we often apply the
most strict selection critera on what to observe there.  Therefore the data
from this telescope make an excellent testbed for our selection method.
However, at other telescopes we observe a broader range of candidates 
to ensure that we are not missing unusual SNe Ia.

It is also important to note that the SN photo-z is only used to
prioritize follow-up observations -- for all scientific studies we use
the measured spectroscopic redshift.  The SN photo-z will be
described in more detail in \citet{2005sullivan}, and the results of
its implementation on our SN Ia confirmation rate are given in
Section~\ref{photoz}.

\section{Observational Techniques}
The observational setup and conditions for each candidate are given in
Table~\ref{obs-table}.  In this section we discuss the observational modes
used in greater detail.

\subsection{Instrument Setup}
All supernovae were observed with the Gemini Multi Object
Spectrographs (GMOS) \citep{2004PASP..116..425H}.  Longslit spectra of the SN
candidates were taken with the GMOS R400 grating (400 lines per
mm), using the 0.75\arcsec\ slit, giving a resolution of 6.9 \AA .  
The detector binning was
set at $2 \times 2$, giving a spatial scale of 0.145\arcsec\ 
per pixel and a spectral dispersion of 1.34 \AA\ per 
binned pixel.  The slit angle was chosen to include the host galaxy wherever 
possible, as the galaxy features provide the most accurate 
determination of the redshift.  

The central wavelength used 
was based on our estimated redshift of each candidate.  
For candidates at estimated redshifts $z \leq 0.4$
we used a central wavelength of 680 nm (with the order blocking filter
GG455), covering 4650\AA\ to 8900\AA .  For higher 
redshifts the 720 nm setting (with the OG515 order blocker) 
was used, giving coverage from 5100\AA\ to 9300\AA .

\subsection{Observing Mode}
The observations were executed in queue mode which allows us to
specify the desired observing conditions.  We require better than
0.75\arcsec\ image quality (corrected to zenith) and clear sky and generally
those requirements were met.  

Observations were made in either nod-and-shuffle (N\&S), electronic
nod-and-shuffle, or classical observing modes.  During a classical
observation, the object and the nearby sky are simultaneously imaged
on adjacent portions of the CCD.  The sky is fit column-by-column and
subtracted, but this can leave systematic residuals near bright and
spatially variable sky lines.  Since this technique requires no additional
overhead we use it on the brightest targets ($\rm \iprime \leq 23.3$).
(All SNLS magnitudes reported in this paper are in the AB system.)

The nod-and-shuffle mode \citep{2001PASP..113..197G} enables more
accurate sky subtraction by nodding frequently (every 60 seconds)
between two positions along the slit.  The charge on the CCD detectors
is simultaneously shuffled between the illuminated science region of
the CCD and the unilluminated (storage) region.  Thus, both the
object and its immediate sky background are imaged on both the same
pixels and at the same slit position.  Systematics due to pixel
response, fringing in the sky lines, slit irregularities, and any
temporal sky variations can be removed by subtracting the N\&S
acquired sky spectrum from the object spectrum.  This method works
well towards the red end of the spectrum (where sky lines are more
problematic) and we found it particularly useful for candidates at
magnitudes fainter than $\rm \iprime > 23.3$.  Typical nod distances
for this method are a few arcseconds.

There are a few drawbacks to nod-and-shuffle.  First, the noise in
sky-subtracted N\&S images is higher by a factor of $\sqrt{2}$
compared to classically-reduced long-slit spectra, because one image is
subtracted from another.  Second, there are increased overheads on any
observation, as each nod cycle adds approximately 24 seconds of
nod time.  For a typical 1800s observation, with 15 (60 second)
nod cycles, these overheads add an extra 360s.  The extra overhead
time can be minimized by choosing a small nod distance, or by
employing the electronic N\&S mode.  The GMOS instrument requires the
use of an On-Instrument Wavefront Sensor (OIWFS), which provides
``fast guiding'' (image motion compensation) and higher order
corrections.  During a normal N\&S observation, this sensor physically
moves during each nod cycle.  The electronic N\&S mode avoids this by
electronically changing the position that the OIWFS guides around
instead of physically moving the sensor.  This decreases the overheads
for a N\&S observation by nearly 200s (for an 1800s exposure), but is
only available for small nod distances (up to 2\arcsec ).  These
small nod distances are sometimes not possible when the candidate
resides in an extended host galaxy.

\section{Data Reduction}
Data reduction was done by collaborators at both University of Toronto
and Oxford University using independent pipelines primarily written in
\textsc{iraf} using the \textsc{gemini} software package version 1.6.
Here we describe the Toronto pipeline, although the reduction methods
are similar for both pipelines.  The independent reductions have been
used to check the consistency of the final output spectra.

In GMOS the spectra are spread over 3 CCDs.  First, bias subtraction
is performed using a master bias for each chip generated from bias
frames covering an entire GMOS queue run.  Next, flat-fielding is done
on a chip-by-chip basis using flat-fields taken before and after each
observation.  The pipeline then branches depending on whether a
nod-and-shuffle or classical observing setup was used.

A side effect of the N\&S observing is that charge-traps, local
defects in the GMOS detectors, appear as low-level horizontal stripes
in the science observations. These are removed using a special N\&S
dark-frame observation.  The dark image reveals the charge-traps, and
is used to construct a bad pixel mask (BPM) that screens out these
areas during image combination.  Additionally, between exposures the 
detector is translated by a few pixels to dilute the
effect of the charge-traps on any one part of the CCD.

The next stage (for both observing modes) is to locate cosmic rays 
in the science frames and to add them to the BPM generated from 
the N\&S dark frame. We use the \textsc{lacosmic} 
package \citep{2001PASP..113.1420V}, 
which locates $>$99\% of cosmic-rays in an image.

Next, N\&S spectra are sky-subtracted using the {\tt gnsskysub}
\textsc{iraf} routine, while classical spectra are sky-subtracted using a
spline function fit along the spatial direction with the science
object pixels excluded from the fit. The resulting frames for
each object are combined using an average combine, rejecting
charge-traps and cosmic rays identified in the BPMs. At this stage a
sky frame median-filtered in the spatial direction is added back to
the data to ensure the correct variance weighting can be used in the
extraction stage. This leaves cosmetically clean 2-D spectral frames.

\begin{figure}
\plotone{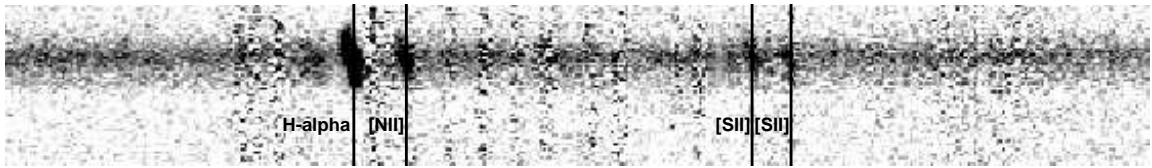}
\caption{Section of a two-dimensional nod and shuffle sky-subtracted
  spectrum.  The wavelength solution is applied to the 2D spectrum and
  the positions of common galaxy lines are overplotted.  Wavelength
  increases to the right; the spatial direction is vertical.  The
  supernova is blended with the host
  galaxy, but the galaxy emission lines are apparent as the dark
  diagonal lines.  The galaxy lines are slanted due to the rotation
  of the galaxy.  Note the excellent sky subtraction -- the subtracted
  sky lines are the vertical bands of increased poisson noise, but
  lack systematic residuals.\label{gallines}}
\end{figure}

The spectra are then extracted using the {\tt apall} task with
variance weighting (using an appropriate effective gain and
readnoise), subtracting off the median sky added back in the previous
step. This produces an error spectrum and a science spectrum
essentially free from systematic sky subtraction errors, ensuring that
careful $\chi^2$ fits can be made to the extracted spectra (discussed
in Section~\ref{chi2}). Wavelength
calibration is then done on the extracted 1-D spectrum using a
solution derived from arc lamps taken once per run, and tweaked using
the night-sky lines for each observation.  The wavelength solution is
then applied to the sky-subtracted 2-D data so that galaxy lines may
be overplotted (Fig.~\ref{gallines}).

The final steps are flux calibration on the extracted 1-D spectrum,
combined with a telluric correction (based on standard star spectra),
and atmospheric extinction correction derived from the
effective airmass of observation and the Mauna Kea extinction curve of
\citet{1987PASP...99..887K}.  The spectra from the 3 chips
are then combined into one spectrum, and an error spectrum is
simultaneously created.  The chip gaps are effectively weighted to 
zero so that they have no influence on the $\chi^2$ fitting.

\section{SN classification}
When possible, the slit was placed through both the SN and the center
of the host galaxy.  In these cases the redshift was determined from
lines in the host galaxy spectrum.  Even if the host and the object
are completely blended, it is still possible to identify the host
galaxy lines as they are narrower than SN features.  Sky-subtracted,
wavelength calibrated, combined two-dimensional spectra were created
for each candidate and the positions of common galaxy spectral
features were overplotted, as shown in Figure \ref{gallines}.  The
lines used to identify the redshift of each host galaxy are identified
in the last column of Table~\ref{der-table}.  In some cases the host
galaxy was too faint and the redshift was determined from
the SN spectrum.

Type Ia SNe can be identified by a lack of hydrogen in their 
spectra, combined with broad (several thousand km s$^{-1}$) 
P-Cyngi lines of elements such as \ion{Si}{2}, \ion{S}{2},
\ion{Ca}{2}, \ion{Mg}{2}, blends of Fe-peak lines, and 
sometimes \ion{Ti}{2}, \ion{O}{1}, or \ion{Fe}{3}.  See 
\citet{1997ARA&A..35..309F} for a review of SN classification,
and \citet{2005hook}, \citet{2005A&A...430..843L}, \citet{2005AJ....129.2352M}, and
\citet{2000ApJ...544L.111C} for particular issues and techniques 
associated with classification of high-redshift SNe.

\subsection{$\chi^2 fitting$\label{chi2}}
High-redshift SN spectra are often blended with their host galaxies, so
determining the SN type can be a challenge (at high redshift the
galaxies have a smaller angular size, so a greater fraction of the
host light is covered by the SN point-spread function).  Some studies
do not attempt to separate SN and host galaxy light, but use a
cross-correlation technique to identify SNe
\citep{2005AJ....129.2352M}.  Other authors attempt to separate SN
and host galaxy light using point source deconvolution
\citep{2005A&A...431..757B}.
Here we use a $\chi^2$ fitting technique to separate SN from host
galaxy light and determine the SN type.  It was first developed by 
\citet{2002AAS...201.9103H} and subsequently used by 
\citet{2005A&A...430..843L} and \citet{2005hook}.   

Each SN spectrum was fit using a $\chi^2$ matching program
which compares the observed spectrum to a library of
template SNe of all types covering a range of epochs.  The redshift, 
amount of host galaxy contamination, and reddening are varied to 
find the best fit.  At a given redshift, the code computes:
$$\chi^2=\sum{\frac{[O(\lambda)-aT(\lambda;z)10^{cA_{\lambda}}-bG(\lambda;z)]^2}{\sigma(\lambda)^2}},$$
where $O$ is the observed spectrum,
$T$ is the SN template spectrum, $G$ is the host galaxy template
spectrum, $A_{\lambda}$ is the redding law, $\sigma$ is the error on
the spectrum, and $a$, $b$, and $c$ are constants that are varied to
find the best fit in host galaxy, template SN, and reddening space.  
If the redshift was not fixed, this equation is then reevaluated over 
a range of redshifts to find the minimum $\chi^2$ in redshift space.  
We use the reddening law of \citet{1989ApJ...345..245C} and $R_V=3.1$.

If the host galaxy spectrum
free of SN light could be extracted, then it was used as the galaxy
spectrum subtracted in the fitting process.  If no galaxy
contamination was evident, then no host was subtracted in the fit.
For other cases, template galaxies from \citet{1996ApJ...467...38K}
and \citet{1997A&A...326..950F} were subtracted.  In cases where
the Hubble type could be estimated from imaging, the galaxy spectral
energy distribution, or narrow galaxy lines, the galaxy type was 
restricted in the fitting procedure.  The SN template library includes
184 SN Ia spectra, 75 SN Ib/c spectra, and 47 SN II spectra.  The
spectra were chosen to have good S/N, to span the widest possible
range of epochs, to have the widest possible wavelength coverage, and to
represent all known SN subtypes. 

\begin{figure}
\plotone{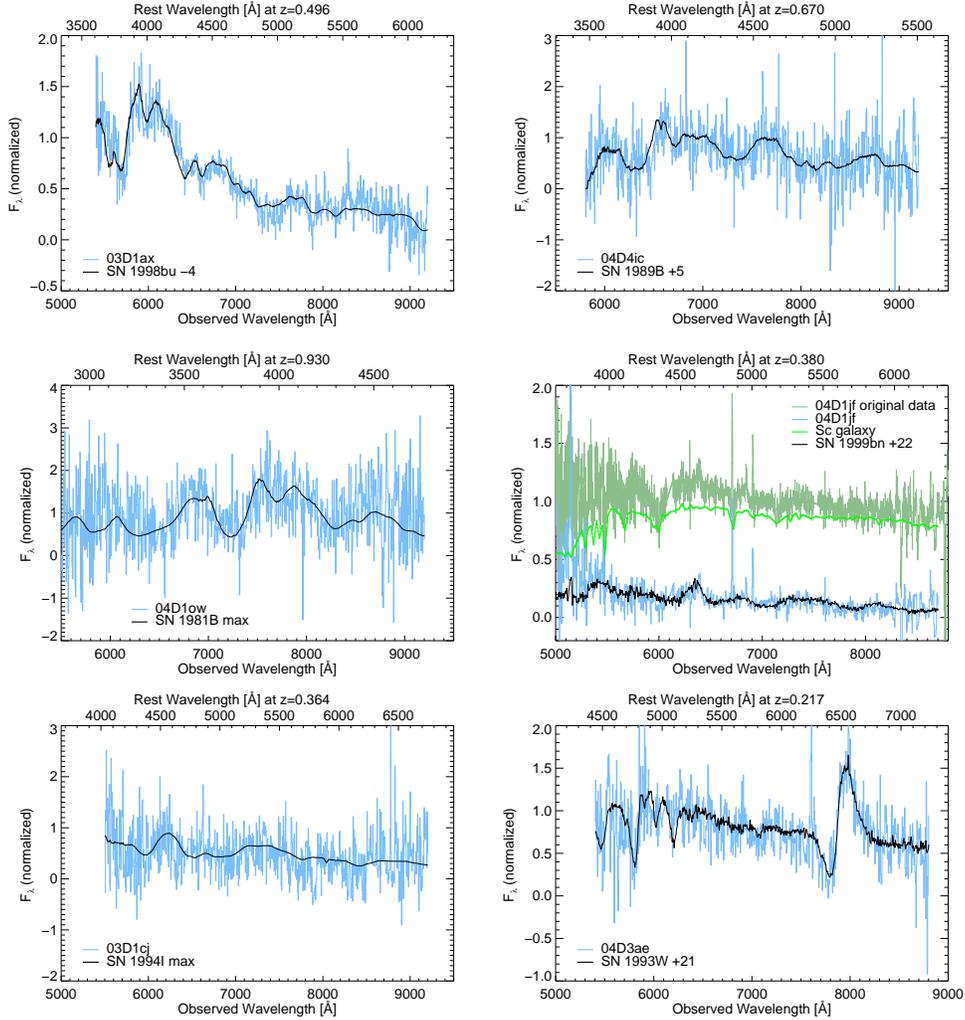}

\vskip -1.8 in
\caption{
  Spectra of SNLS
  candidates observed by Gemini.  One spectrum from each
  SN Ia confidence index is shown as an example.  All other spectra
  are available online.  When there is only minor galaxy
  subtraction, only the subtracted spectrum (rebinned to 5\AA )is shown.  In cases where
  the galaxy subtraction is important, the original (unbinned) spectrum, the
  galaxy subtracted spectrum and the smoothed host galaxy that was 
  subtracted are shown.  Top-left: 5 (Certain Ia), the Si
  4000\AA\ feature is obvious;
  top-right: 4 (Highly probable Ia), all features match, but at this phase 
  the Si 4000\AA\ feature is not a definitive detection; middle-left: 
  3 (Probable Ia) an SN Ia is the best fit, but the lower S/N at high
  redshift leaves some room for doubt;
  middle-right: 2 (Unidentified), significant host contamination,
  combined with the fact that many SN types look similar at a late
  phase means that this SN type is uncertain, despite the match to an
  SN Ia shown; bottom-left: 1 (Probably not a Ia) An SN Ic is a good
  fit to the spectrum, but the redshift is uncertain and the S/N is
  too low to make a definitive classification;
  bottom-right: 0 (Certainly not a Ia), in this case a Type II.  
  For Index 5, 4, 3, 1, and 0 the gray line (light blue online) 
  shows the data after host galaxy subtraction (if necessary),
  rebinned to 5\AA .  For Index 2, the top gray line (dark green) 
  shows the original data, overplotted with the best-fit
  host galaxy template (bold lighter green).  The lower gray line
  (light blue) shows the data after
  host subtraction and rebinning --- the SN type could not be identified.
  \label{specfig}}
\epsscale{1.0}
\end{figure}

The $\chi^2$ fitting program produces a list of the best matching SN
templates, host galaxy templates, redshifts, and reddening.  As in SN
detection, it is not yet possible to fully automate this process -- a
human must still inspect the results and make the ultimate
determination of the SN type.  The type of the SN is estimated and
placed into one of six categories reflecting the type and uncertainty
in the classification (Section \ref{CI}; Fig. \ref{specfig}).  The
spectroscopic epoch is determined from the weighted mean of the best 5
epochs (weighted by the $\chi^2$ of each fit).  The average of several
epochs was chosen to dilute the effect of outliers, and to help smooth
out the diversity in SN Ia spectra.  The exact number of epochs
averaged has little effect on the result, since the average is
weighted by the $\chi^2$ of each fit --- the best few fits will be the
dominant contributors to the average.  We find $\chi^2=1$ for an error
of $\sigma=2.5$ days on the spectroscopic
date determination."

\subsection{SN Ia confidence index\label{CI}}

Two factors make it harder to classify SNe at high redshift as
compared to their counterparts at low redshift: they generally have
lower signal-to-noise spectra and there is a greater degree of
contamination from the host galaxy.  While many classifications are obvious,
others have a degree of uncertainty associated with them.  To quantify
this, after examination of its spectrum we give each SN a
SN Ia confidence index:

\begin{enumerate}
\item[5] {\bf Certain Ia:} The spectrum shows distinctive features of
  an SN Ia such as \ion{Si}{2} or \ion{S}{2}.  Often \SiII\ 6150\AA\ 
  and \ion{S}{2} 5400\AA\ are redshifted out of the observed spectral
  range, so \ion{Si}{2} 4000\AA\ is used as the key indictor for SNe
  Ia.  However, at some phases, or for some SN Ia subtypes (\eg\ 
  SN~1991T), \ion{Si}{2} 4000\AA\ would not be expected.  In these
  cases the candidate can be classified as a category 5 if the
  spectrum is an exact match to the overall spectral energy
  distribution (SED) of an SN Ia (at the phase indicated by the
  lightcurve) and no other type of SN matches.
\item[4] {\bf Highly probable Ia:} The spectrum is a match to a Ia,
  but lacks an unambiguous detection of one of the features that is
  unique to SNe Ia (\SiII\ or \SII ).  Other SN types do not match the
  spectrum well.  These candidates usually have another piece of
  confirming evidence, such as a lightcurve consistent with a SN Ia at
  the measured redshift, or they are found in an E or S0 galaxy.
\item[3] {\bf Probable Ia:} The spectrum matches an SN Ia better than
  any other SN type, but another SN type (usually SNe Ic) is not
  ruled out from the spectrum alone. This is either because the
  spectrum has low S/N or because other SNe look similar at the same
  phase.  These candidates have a lightcurve consistent with an SN Ia
  at the measured redshift, and the spectrum has a phase consistent
  with the lightcurve.  These SNe are denoted as Ia*, following the
  notation of \citet{2005A&A...430..843L}.
\item[2] {\bf Unknown:}  The type cannot be determined from the
  spectrum.  Often the spectrum has low signal-to-noise, or there is 
  too much host galaxy contamination to make a
  reliable determination.
\item[1] {\bf Probably not a Ia:}  The spectrum has features
  marginally inconsistent with an SN Ia, but the type cannot be
  unambiguously determined.  
\item[0] {\bf Not a Ia:}  The spectral features are inconsistent with a
  SN Ia.  In this case the spectrum can usually be identified as an SN
  II, SN Ib/c, or an AGN (though note that there are no clear cases of
  AGN from Gemini spectra --- these were screened out in advance).
\end{enumerate}

The primary means of classification ibfinpstws the SN spectrum, although we
use all information available to supplement this including the
lightcurves, colors, and the agreement between lightcurve phase and
the spectroscopic phase determined from the fitting process.  For
example, if the type cannot be determined from the spectrum it would
normally be classified as index 2.  But if the lightcurves are also
inconsistent with the lightcurve of an SN Ia, then it would be moved
into index 1 or 0.  The host galaxy spectrum is never used on its own
to classify a candidate, but if the host is clearly an elliptical
galaxy, then this information may be used in conjunction with the
candidate spectrum and the lightcurve to solidify the status of a
candidate as a probable SN Ia.  We emphasize that the lightcurve alone
is never used to classify an SN Ia -- for a candidate to have a
classification of SN Ia or SN Ia* it must have a spectrum matching
with the expected spectral energy distribution of an SN Ia at the
phase determined from the lightcurve.

One representative spectrum from each category is shown in 
Figure~\ref{specfig}.  All spectra discussed in this paper are 
available in the online version of this article.  

\section{Results}
\label{results}
The SNLS officially started in June 2003 (there was a presurvey
ramp-up to full operations), and we began using Gemini-N and S to
observe SN candidates in August 2003 (semester 2003B).  We usually
obtain the Gemini data within a day or two of the observations and produce
``real-time'' reductions.  After the full calibration data is released at
the end of a run the data are rereduced.  Here we report on the final
spectroscopic reductions through November 2004.  All of the spectra
used in this study are available in the online version of the paper.
 
Table~\ref{der-table} lists properties derived from the
observations of each candidate, such as type, redshift, and epoch
relative to maximum light.  Table~\ref{class-host-table} shows the
distribution of SN types with respect to host galaxy type.  While it is not
always possible to extract the host galaxy separately and examine its
SED, galaxy lines are apparent because they are narrower than SN
features.  Therefore we group galaxies into absorption-line galaxies
and emission-line galaxies (if a galaxy has any emission lines it is
considered an emission-line galaxy).  Just as at low redshift, in our
sample core-collapse SNe are never seen in absorption-line (early
type) galaxies.

\begin{figure}
\plotone{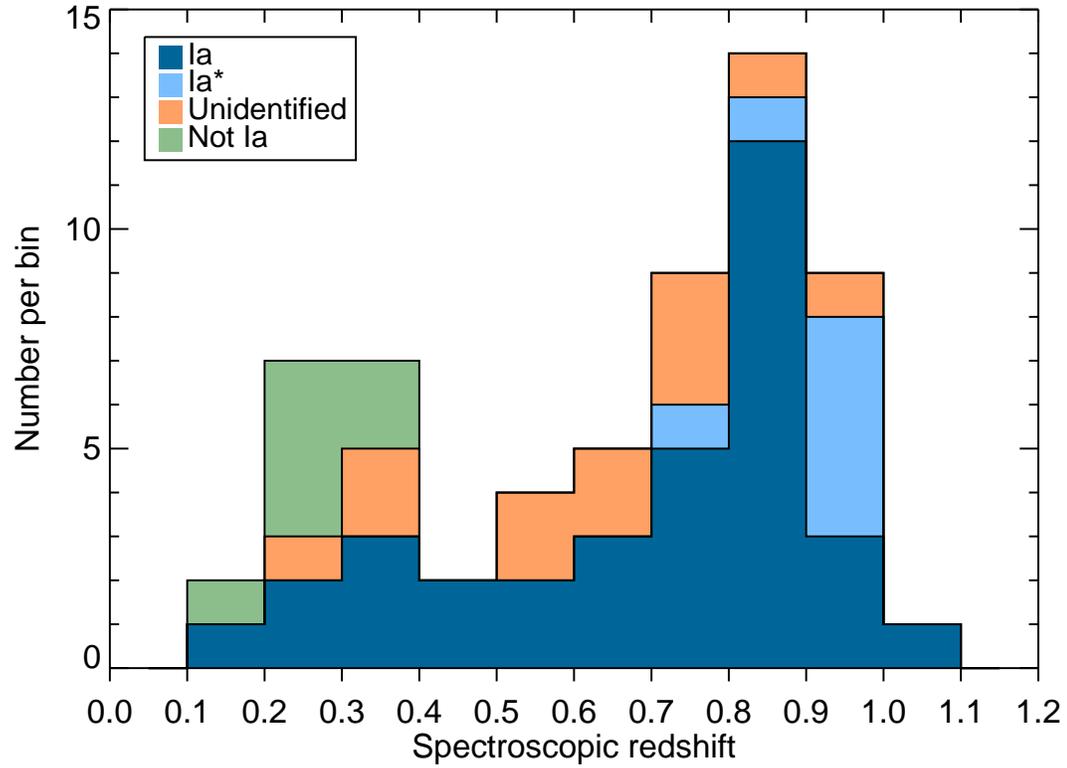}
\caption{Histogram of the redshifts of candidates observed by Gemini,
  where the highest redshift candidates were usually observed.
  Note that SN Ia classifications are less certain (Ia*) at higher
  redshifts where the S/N is lower and scarce restframe UV observations must
  be used to classify the SN.  Non-Ia candidates were typically at a 
  lower redshift than that predicted
  for an SN Ia using the SN photo-z, since core collapse SNe are generally fainter than SNe Ia.  The median redshift of SNe Ia/Ia* is
  0.81. \label{hist}}
\end{figure}

Figure~\ref{hist} is a histogram of the
number of candidates of each different type with redshift.  For
clarity we group index 0 and 1 SNe together as ``Not Ia'' and index 4
and 5 SNe together as ``Ia.''  SNe with weaker classifications (index
3) are plotted separately as ``Ia*.''  Candidates that could not be
identified (index 2), but whose redshifts could be determined, are
also shown.  It is apparent that we targeted the highest redshift
observations at Gemini --- the median redshift of SNe Ia/Ia* is 0.81.
As the redshift increases the fraction of less secure identifications
rises, because for faint SNe it is hard to achieve a high signal-to-noise
ratio in a reasonable integration time.  Furthermore, at the
higher redshifts one must rely more on the rest-frame UV light to
classify the SNe, and template UV observations of all SN types are scarce.

Core collapse SNe cluster at lower redshifts on the histogram because
they are usually intrinsically fainter.  For example, a candidate with
a peak magnitude of $\rm \iprime = 24$ could be an SN Ia at $z=0.9$ or
could be a Type II SN at a lower redshift.

\subsection{SN photo-z results\label{photoz}}
The technique of the SN photo-z is introduced in
Section~\ref{photozintro} and described in much greater detail in
\citet{2005sullivan}.  Here we report the result of its application on
our SN Ia confirmation rate.  The SNLS implemented the SN photo-z and
improved real-time photometry in March 2004, although in two cases
after this (04D1dr and 04D4ft), candidates had to be observed
spectroscopically before a photo-z could be obtained.  When no photo-z
information was available before selecting candidates for
spectroscopy, 14/26 (54\%) of the candidates were confirmed as SNe Ia
--- close to previosly published rates.  However, when we did have an
SN photo-z to guide the decision making, the Ia confirmation rate
jumped to 27/38 (71\%).  There will be always be some unidentified
candidates for which it is difficult to estimate a type from
spectroscopy --- these may still be SNe Ia, but are too buried in a
host, or have a spectrum with too low S/N, to be identified.  However,
where the photo-z excels is in its power to reject candidates that are
not SNe Ia.  With no prior photo-z information, 7/26 (27\%) of
candidates with Gemini spectroscopy were found to be certainly not or
probably not SNe Ia.  After implementation of the technique, only 3/38
(8\%) of the observations were certainly not or probably not SNe Ia.

Another benefit of the SN photo-z is that it provides a prediction of
the time of maximum light for an SN Ia, allowing observations to be
scheduled to within a few days of this date for maximum efficiency and
minimal host galaxy contamination.  Figure~\ref{hiepoch} shows the
distribution of the SNe~Ia observed by Gemini with respect to maximum
light.  Ninety percent of SNe Ia were observed within 0.5 mag of
maximum light, and over half of the SNe Ia were observed within 0.1
mag of maximum light.  It is clear that the flexibility provided by
queue observing plays a large role in optimizing the efficiency of the
spectroscopic classification of targets in the SNLS.  Note that it is
not always possible to schedule observations at maximum light because
we require dark time to observe such faint targets, and GMOS is not
always on the telescope and available in queue mode.

\subsection{Optimal time for spectroscopy}
\begin{figure}
\plotone{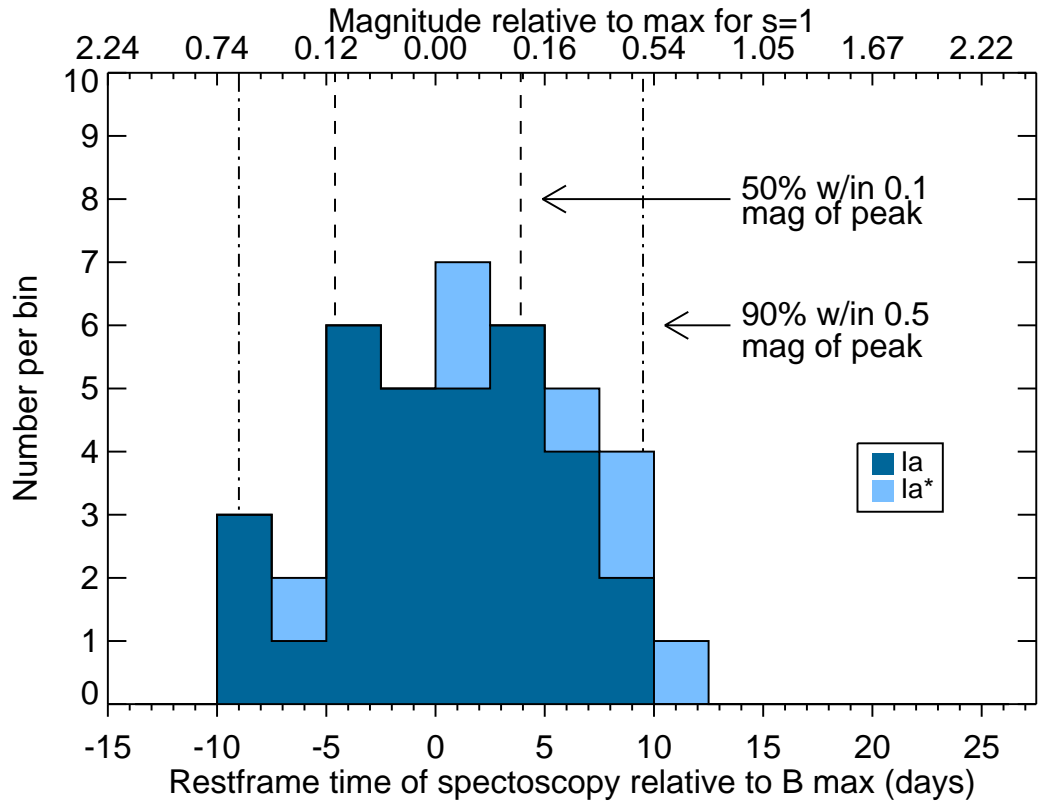}
\caption{Histogram of rest frame epoch relative to restframe B-band maximum
  light for Type Ia/Ia* SNe.  Since we can predict the time of maximum
  light before a candidate is sent for spectroscopy, we can plan the
  observations so that the contrast between the SN and the host
  galaxy is at a maximum.  More than 50\% of SNe were observed
  within 0.1 mag of maximum light, and 90\% of SNe were observed
  within 0.5 mag of maximum.  It is easier to classify SNe Ia with
  certainty if spectra are taken just before maximum light.\label{hiepoch}}
\end{figure}

Figure~\ref{hiepoch} also shows that the least solid classifications
(index 3), labeled Ia*, were usually observed after maximum light.
SNe before maximum were almost always classified with more certainty.
This is partially because after maximum light, especially around +7 to
+10 days after max, it is often difficult to distinguish between SNe
Ia and SNe Ic.  By this time certain distinguishing SN Ia features,
such as \SiII\ 4000\AA , may no longer be apparent, and the SN Ia line
velocities have decreased to the range more typical of those found in
SNe Ic.

The more uncertain classifications (SNe Ia*) often occur near maximum
light, because the most difficult SNe (those at the highest redshift
or with the most host galaxy contamination) can only be observed
near maximum light.  Before or after maximum light they are so faint
that they would not be placed in the spectroscopic observing queue.

An added benefit to obtaining early spectroscopy is
that SNe Ia show the greatest diversity at early 
times \citep{2001ApJ...546..734L}, when the
spectroscopy is probing the outer layers of supernova.
Figure~\ref{hiepoch} shows that one compromise would be to target
observations at about one week before maximum light.  At $-7$d, a typical
Ia is only about 0.25 fainter than at peak, near enough to peak
brightness to make the observations feasible.  At the same time it
would provide an opportunity to observe SNe Ia when they show the
greatest diversity and also are the most distinct from SNe Ic.

This window of opportunity is very narrow, however.  At $-10$d a typical
SN Ia is 0.75 mag fainter than at peak in the restframe B-band ---
still too faint compared to its host galaxy, and by $-4$d SN~Ia
spectra have lost some of their diversity.  

\subsection{Identification in the presence of host contamination}
\begin{figure}
\plotone{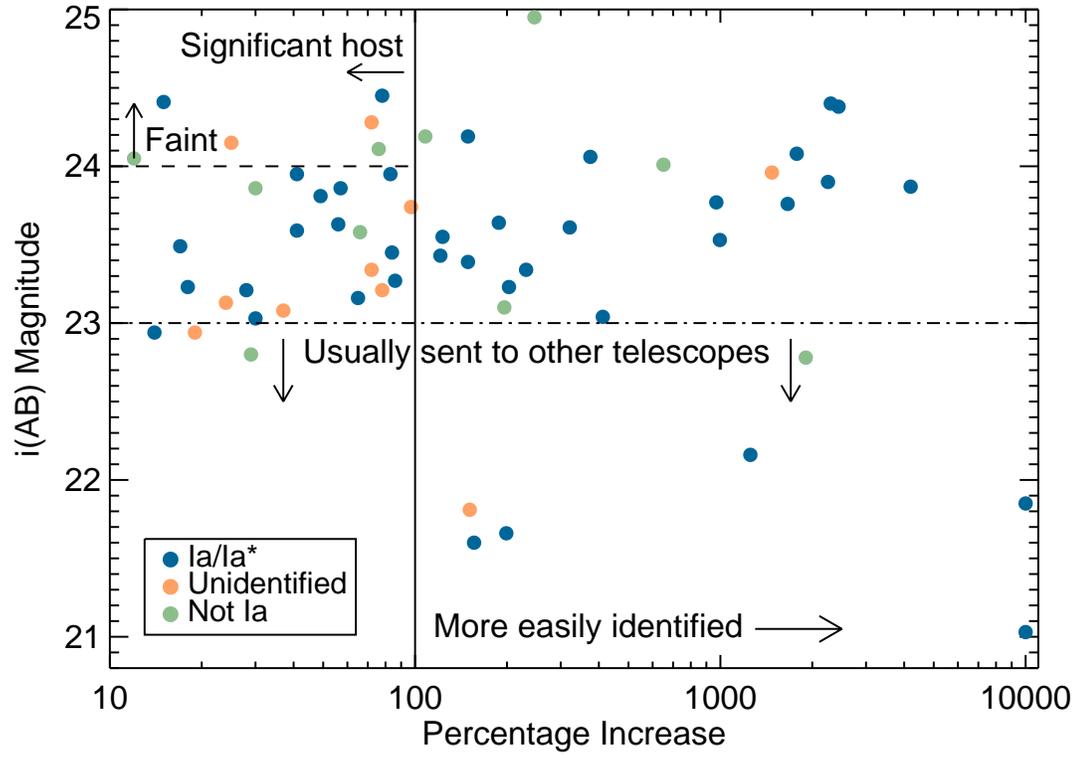}
\caption{\iprime\ Magnitude at time of spectroscopy versus
  percentage increase.  Typically candidates were only sent to Gemini
  if $23 < \rm i\arcmin\ <24.5$.  When the candidate is
  brighter than the host (greater than 100\% increase) it is
  identified 93\% of the time.  
\label{imagpi}
}
\end{figure}

Figure~\ref{imagpi} shows the \iprime\ magnitude at the time of
spectroscopy versus the percentage increase in \iprime\ brightness in a 6
pixel (1.12\arcsec ) diameter at the time of spectroscopy.  The
percentage increase is measured relative to the reference image, where
there is no supernova light.  The \iprime\ magnitude and percentage
increase were measured from CFHT images, interpolated to the time of
spectroscopy.  Typically candidates were sent to Gemini only if they were in the
magnitude range $23 < \rm i\arcmin\ <24.5$.  (There are some
exceptions, especially in the D3 field, which cannot be seen by the VLT.)  

It is apparent from Figure~\ref{imagpi} that when the SN signal is
greater than that of the host galaxy, candidate identification 
is relatively easy --- when the percentage increase is greater 
than 100\% the candidates were not identified only 7\% (2/30) of the time.

\begin{figure}
\plotone{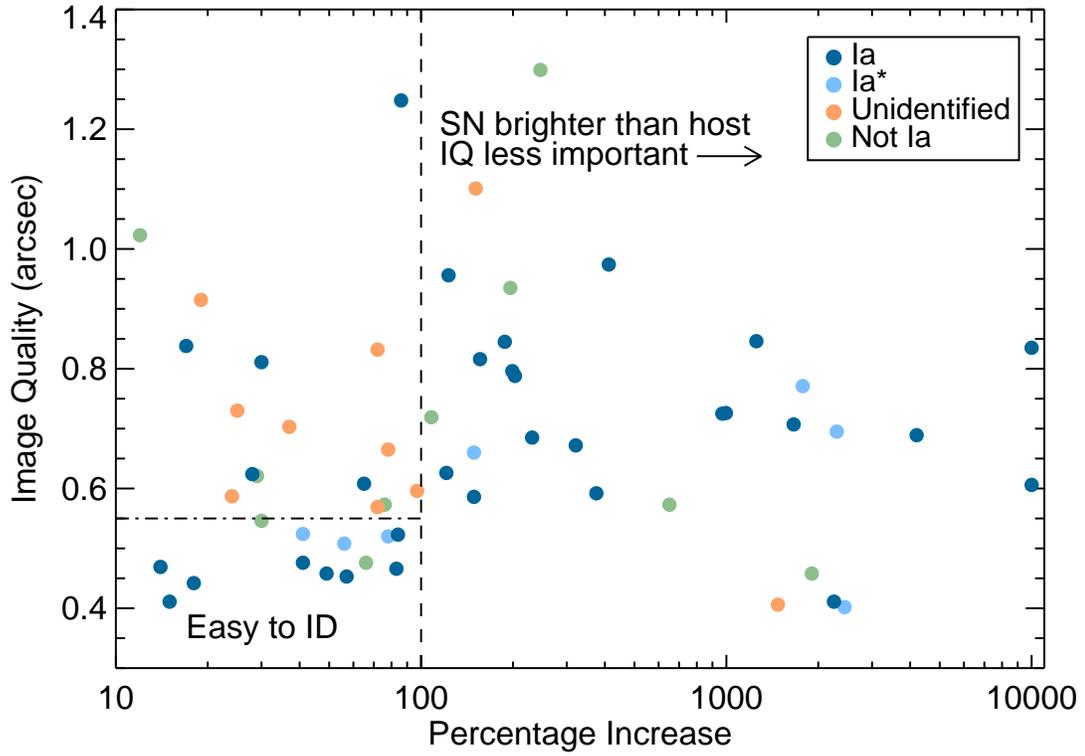}
\caption{Image quality (IQ) at the time of spectroscopy versus 
  percentage increase at the time of spectroscopy.  If the SN was
  brighter than the host (\% increase $>$ 100), IQ plays a less
  significant role in determining whether a candidate is identified.
 However, candidates can still be identified in the presence of
 significant host contamination if the seeing is exceptional.
  If the IQ was less than 0.55$\arcsec $ candidates were identified
  88\% of the time (only SNe with both a percentage increase
  and IQ measurement are plotted in the figure).  \label{iqpi}}
\end{figure}

Clearly, image quality (IQ) plays a role in whether or not SNe can be
successfully identified in the presence of significant host galaxy
light, as shown in Figure~\ref{iqpi}.  We plot the image quality at 
the time of spectroscopy (determined from the Gemini
acquisition image) against the percent increase (determined from CFHT
images as described above).  If the IQ was better than 0.55$\arcsec $,
candidates were identified 88\% of the time.  In good 
seeing the SN light is more concentrated and can be extracted 
in a narrow aperture, even in the presence of host contamination.  

\subsection{Success of $\chi^2$ fitting}
The $\chi^2$ matching of the SN spectrum against a host and SN 
spectral template library produces more identifications, more robust
identifications, and greater coverage of parameter space than
traditional methods (unaided expert matching by eye).  This method
works optimally with spectra that are free from systematic deviations,
and whose errors are well characterized.  It works exceptionally well
with GMOS nod and shuffle data, where systematic effects associated
with sky subtraction are almost completely removed.  Correcting the spectra for
telluric features also helps, as does a carefully generated error
spectrum.  

While both the spectra in this paper and in
\citet{2005A&A...430..843L} were classified by the same person (DAH), here the
$\chi^2$ matching software is upgraded with more template supernova
and galaxy spectra.  \citet{2005A&A...430..843L} were not able to
identify any candidates with a percentage increase below 25\%, but
in this work we have successfully identified several candidates with percent
increases between 15\%--25\%.  It is unclear if this is
due to software improvements, observing conditions, differences 
in the spectrographs used, or a better selection of candidates.

\begin{figure}
\plotone{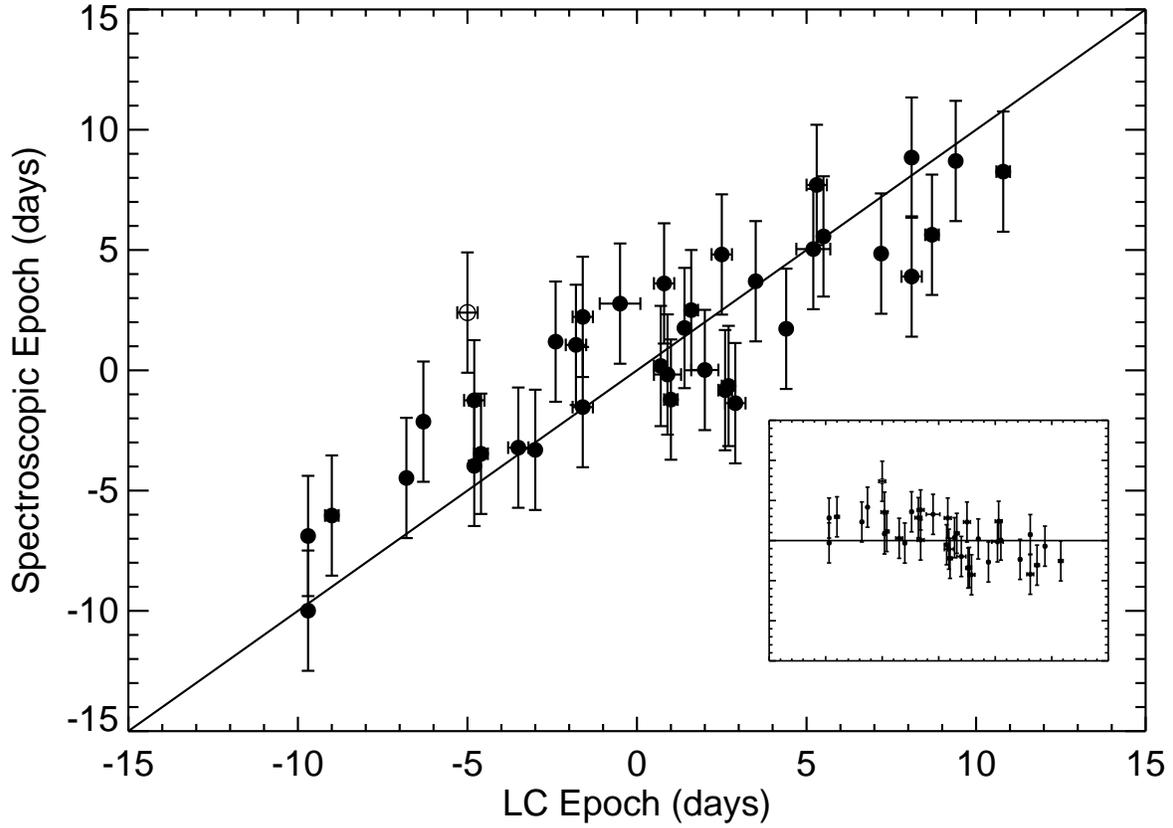}
\caption{Rest frame epoch of spectroscopy relative to rest B-band
  maximum light as determined by fits to the lightcurve and fits to
  the spectra.  The line shows where SNe should lie if the lightcurve
  epoch is equal to the spectroscopic epoch -- it is not a fit to the
  data.  The lightcurve epoch is from fits to the final reductions of
  the lightcurve data where possible, but in some cases it is from
  fits to preliminary (real-time) data.  The open circle
is a spectroscopically peculiar (SN 2001ay-like) SN 
for which there are few templates in the spectroscopic matching library.
The inset shows the difference between the spectroscopic epoch and the
lightcurve epoch as a function of lightcurve epoch, with the same
units as the larger graph.
Note that from -4d to +4d, SN Ia spectra are similar, so 
an automatic determination of a date from fits to the 
spectrum can be difficult.  There is also a tendency to overpredict
the spectroscopic phase at early times.  This is probably due to 
the scarcity of early SN Ia spectra.
\label{date}}
\end{figure}

Figure~\ref{date} also shows that the spectroscopic epoch determined
by the fitting program matches well with the epoch determined from the
lightcurve.  Unlike the spectral feature age technique
\citep{1997AJ....114..722R}, it is important to note that this program
is not specifically tuned to determine the epoch of an SN Ia, and it
makes no assumptions that the input spectrum is a Ia (it is possible
for it to determine the epoch of an SN Ib/c for example, although this
has not been extensively tested).  We find the remarkable 
result that our spectroscopic fitting technique can determine the
epoch to within 2.5 days, despite the low S/N and significant 
host contamination in this data set.

\section{Discussion} 
Before comparing these confirmation rates to previous work, a few 
caveats are necessary.  First, not all spectroscopic follow-up programs
have as their goal the confirmation of candidates as SNe Ia.  One may
wish to determine the types of all transients to have a better
determination of SN rates.  Or in some cases the goal may be to
spectroscopically identify SNe II, as they also have cosmological
utility.  While we have pursued these goals at other telescopes, the
role of Gemini has been largely to confirm SNe Ia.  Second, different
programs may have different criteria for what is called an SN Ia/Ia*.
For example, we have lightcurve infomation available at the time of 
classification, which was not always the case in previous studies.
Still, our criteria are as close as possible to the criteria used in
other published works.  We adopt similar criteria to that of  
\citet{2005AJ....129.2352M}, and we are using the
the same human classifier (DAH) and classification program (albeit 
slightly upgraded) as \citet{2005A&A...430..843L}.

Despite these caveats, it is important to have some comparison
to previous work.
A comparison to \citet{2005A&A...430..843L} is problematic because of the 
inhomogeneous nature of the sample and techniques presented there. 
They present results from six separate searches, some classical, and
some rolling, some targeting $z\sim0.5$, and some targeting $z>1$.


The most appropriate comparison to this work is to consider all
searches in Lidman et al. that targeted $z<1$.  In this case the 
SN Ia/Ia* confirmation rate was 62\%  (median SN Ia $z=0.51$).  
A similar SN Ia confirmation rate is found if
the Lidman et al. search 4 (the rolling search) is excluded, but 
SNe followed by the SCP at all telescopes for $z<1$ are included.

Since rolling searches find all SNe above some magnitude cutoff, the
percentage of SNe Ia in the sample is lower than in a classical
search.  In their first two years of operation, ESSENCE 
\citep{2005AJ....129.2352M}, had a 43\% SN Ia/Ia?
confirmation rate with a median redshift of 0.43 [though in the 
third year the SN Ia/Ia? confirmation rate rose to $\sim$60\%
(Matheson, private communication)].  Again, a direct
comparison to the results presented here may not be appropriate if the two
searches have different goals or different criteria for counting SNe
Ia, but does make the point that if the goal is to maximize the yield
of SNe~Ia, a method like the SN photo-z is helpful.

Despite the difficulties in comparing this study to previous ones as
noted above, we can draw some broad conclusions from the results.
First, to make the most efficient use of 8--10m time it is helpful to
have as much information as possible before sending a target to the
telescope.  In a rolling search, more information is available at
follow-up time than in a classical search, so new techniques are
called for.  Here we have demonstrated that by fitting the available
data before spectroscopy we can improve the SN Ia/Ia* confirmation
rate over our rates when the technique was not applied.  Our SN Ia/Ia*
confirmation rate (71\%) is also an increase over comparable
previously published rates (43\%-62\%), despite being at a much higher
median redshift (z=0.81 vs. z=0.4-0.5).  By making it possible to
separate likely SNe Ia from likely core collapse SNe before a large
amount of telescope time is invested in spectroscopy, the techniques
demonstrated here improve the success rate of all follow-up programs,
no matter their goal.

Over the five-year lifetime of the SNLS, the these techniques 
will result in $\sim 170$ more confirmed SNe Ia.  We aim to 
spectroscopically observe $\sim 1000$ SN candidates during the survey.
In the absence of the photo-z this would result in $\sim$ 540 SNe Ia,
but with the photo-z we could achieve 710 SNe Ia.
This is equivalent to adding another year to the survey.


\section{Conclusions}
We have demonstrated several techniques for improving the yield of
spectroscopically identified SNe Ia at high redshift ($0.3 \leq z \leq
1.0$).  The SN
photo-z is shown to effectively screen out non-Ia candidates before
they are observed spectroscopically.  When no photometric redshifts
were available, 35\% of candidates turned out to be probably or
certainly not SNe Ia after spectroscopy (CI 0 or 1).  With photo-z information,
the non-Ia ``contamination'' rate dropped to 8\%.  Using the photo-z
we show that we can schedule SN observations to within a few tenths of
a magnitude of maximum light, and that the optimal phase for SN Ia
identifcation and diversity is $-7$ days.  After
spectroscopy, we show that by $\chi^2$ fitting of template SNe we can
effectively subtract host light and determine the type of an SN when
the SN is only 15\% as bright as the host in some cases.  
For targets
where the SN is at least as bright as the underlying host, or
when the image quality is exceptional (better than 0.55\arcsec ), the 
candidate is identified more than 90\% of the time.  
Using $\chi^2$ fitting we can
also obtain an independent measurement of the spectroscopic epoch that
agrees well with the phase determined from the lightcurve.  

These techniques have been developed using the first year's
observations of the highest redshift candidates of the SNLS at 
Gemini North and South.  Of the candidates observed at Gemini, 
41/64 are certain or probable SNe Ia.  This is
roughly one-third of the spectroscopic follow-up program of the SNLS.
The techniques outlined here will add $\sim 170$ more confirmed SNe Ia
over the five-year project to discover,
confirm, and follow $\sim 700$ SNe Ia to measure the equation of state
of Dark Energy.
\acknowledgements

The SNLS collaboration gratefully acknowledges the assistance of Pierre
Martin and the CFHT Queued Service Observations team.  Jean-Charles
Cuillandre and Kanoa Withington were also indispensable in making
possible real-time data reduction at CFHT.  We also thank Gemini queue
observers and support staff, especially Inger J{\o}rgensen, Kathy Roth,
Percy Gomez, and Marcel Bergmann, for both taking the data presented
in this paper and making observations available quickly.
Canadian collaboration members acknowledge support from NSERC and
CIAR; French collaboration members from CNRS/IN2P3, CNRS/INSU and CEA;
Portuguese Collaboration members acknowledge support from
FCT-Funda\c{c}\~ao para a Ci\^encia e Tecnologia.

SNLS relies on observations with MegaCam, a joint project of
CFHT and CEA/DAPNIA, at the Canada-France-Hawaii Telescope (CFHT)
which is operated by the National Research Council (NRC) of Canada, the
Institut National des Science de l'Univers of the Centre National de la
Recherche Scientifique (CNRS) of France, and the University of Hawaii. This
work is based in part on data products produced at the Canadian
Astronomy Data Centre as part of the Canada-France-Hawaii Telescope Legacy
Survey, a collaborative project of the National Research Council of
Canada and the French Centre national de la recherche scientifique.

This work is also based on observations obtained at the Gemini
Observatory, which is operated by the Association of Universities for
Research in Astronomy, Inc., under a cooperative agreement with the
NSF on behalf of the Gemini partnership: the National Science
Foundation (United States), the Particle Physics and Astronomy
Research Council (United Kingdom), the National Research Council
(Canada), CONICYT (Chile), the Australian Research Council
(Australia), CNPq (Brazil) and CONICET (Argentina).  This research
used observations from Gemini program numbers: GN-2004B-Q-16,
GS-2004B-Q-31, GN-2004A-Q-19, GS-2004A-Q-11, GN-2003B-Q-9, and
GS-2003B-Q-8.



\clearpage


\begin{deluxetable}{llllrclrrr}
\tablewidth{0pt}
\tablecaption{SNLS SN Candidates Observed with Gemini\label{obs-table}}
\tablehead{
\colhead{SN} &\colhead{RA (2000)} & \colhead{Dec (2000)} &
\colhead{UT Date} & \colhead{Exp.\tablenotemark{a}} & \colhead{Mode\tablenotemark{b}} &
\colhead{$\lambda_{\rm c}$\tablenotemark{c}} & \colhead{IQ\tablenotemark{d}} &
\colhead{Mag\tablenotemark{e}}& \colhead{\%I\tablenotemark{f}}
}
\startdata
03D1as & 02:24:24.520 & -04:21:40.19 & 2003-09-27 & 6000& N+S & 720 & 0.41 & 23.96 & 1475 \\ 
03D1ax & 02:24:23.320 & -04:43:14.41 & 2003-09-29 & 2400& C & 720 & 0.61 & 23.16 & 65 \\ 
03D1bk & 02:26:27.410 & -04:32:11.99 & 2003-09-28 & 4800& N+S & 720 & 0.47 & 23.95 & 83 \\ 
03D1cj & 02:26:25.081 & -04:12:39.89 & 2003-10-26 & 5400& N+S & 720 & 0.57 & 24.11 & 76 \\ 
03D1cm & 02:24:55.288 & -04:23:03.68 & 2003-10-27 & 5400& N+S & 720 & 0.72 & 23.77 & 970 \\ 
03D1co & 02:26:16.238 & -04:56:05.76 & 2003-11-01 & 7200& N+S & 720 & 0.84 & 23.64 & 188 \\ 
03D1ew & 02:24:14.088 & -04:39:56.98 & 2003-12-21 & 7200& N+S & 720 & 0.71 & 23.76 & 1661 \\ 
03D1fp & 02:26:03.073 & -04:08:02.02 & 2003-12-26 & 7200& N+S & 720 & 0.55 & 23.86 & 30 \\ 
03D1fq & 02:26:55.683 & -04:18:08.10 & 2003-12-24 & 5400& N+S & 720 & 0.48 & 23.59 & 41 \\ 
03D4cj & 22:16:06.660 & -17:42:16.72 & 2003-08-26 & 2700& C & 680 & 0.83 & 21.85 & 10000 \\ 
03D4ck & 22:15:08.910 & -17:56:02.17 & 2003-08-27 & 2400& C & 680 & 0.46 & 22.78 & 1905 \\ 
03D4cn & 22:16:34.600 & -17:16:13.55 & 2003-08-27 & 4800& C & 720 & 0.46 & 23.81 & 49 \\ 
03D4cy & 22:13:40.460 & -17:40:53.90 & 2003-09-26 & 5400& N+S & 720 & 0.59 & 24.19 & 149 \\ 
03D4cz & 22:16:41.870 & -17:55:34.54 & 2003-09-27 & 3600& C & 720 & 0.41 & 24.41 & 15 \\ 
03D4fd & 22:16:14.471 & -17:23:44.37 & 2003-10-24 & 3600& N+S & 720 & 0.67 & 23.61 & 321 \\ 
03D4fe & 22:16:08.844 & -17:55:19.21 & 2003-10-24 & 3600& N+S & 720 & 0.48 & 23.58 & 66 \\ 
03D4gl & 22:14:44.177 & -17:31:44.47 & 2003-10-29 & 3600& N+S & 720 & 0.63 & 23.43 & 121 \\ 
04D1de & 02:26:35.925 & -04:25:21.65 & 2004-08-17 & 7200& N+S & 720 & 0.51 & 23.63 & 56 \\ 
04D1dr & 02:27:23.905 & -04:51:27.43 & 2004-08-14 & 5400& N+S & 720 & 0.57 & 24.28 & 72 \\ 
04D1hd & 02:26:08.850 & -04:06:35.22 & 2004-09-13 & 2400& C & 680 & 0.85 & 22.16 & 1254 \\ 
04D1ho & 02:24:44.856 & -04:39:15.55 & 2004-09-16 & 3600& C & 720 & 0.59 & 23.13 & 24 \\ 
04D1hy & 02:24:08.678 & -04:49:52.22 & 2004-09-11 & 5400& N+S & 720 & 0.73 & 23.53 & 996 \\ 
04D1jf & 02:25:18.914 & -04:49:09.05 & 2004-10-13 & 2400& C & 680 & 0.92 & 22.94 & 19 \\ 
04D1ln & 02:25:53.482 & -04:27:03.75 & 2004-10-17 & 2400& C & 680 & 0.62 & 22.80 & 29 \\ 
04D1ow & 02:26:42.708 & -04:18:22.55 & 2004-11-08 & 5400& N+S & 720 & 0.69 & 23.87 & 4201 \\ 
04D2aa\tablenotemark{g} & 10:02:02.100 & +02:40:51.76 & 2004-01-23 & 7320& N+S & 720 & 0.58 & 24.19 & \nodata \\ 
04D2ad\tablenotemark{g} & 10:00:08.093 & +02:39:01.40 & 2004-01-22 & 7320& N+S & 720 & 0.54 & 24.17 & \nodata \\ 
04D2ae\tablenotemark{g} & 10:01:52.414 & +02:13:21.11 & 2004-01-21 & 5490& N+S & 720 & 0.75 & 23.64 & \nodata \\ 
04D3aa & 14:16:49.935 & +52:45:31.12 & 2004-01-30 & 5400& N+S & 720 & 1.02 & 24.05 & 12 \\ 
04D3ae & 14:22:21.569 & +52:21:39.21 & 2004-01-25 & 2400& C & 680 & 0.94 & 23.10 & 196 \\ 
04D3ax & 14:22:39.072 & +52:51:52.57 & 2004-01-28 & 5400& N+S & 720 & 1.30 & 24.95 & 246 \\ 
04D3bf & 14:17:45.096 & +52:28:04.31 & 2004-02-17 & 2700& C & 680 & \nodata & 23.16 & \nodata \\ 
04D3dd & 14:17:48.431 & +52:28:14.72 & 2004-04-25 & 5400& N+S & 720 & 0.59 & 24.06 & 375 \\ 
04D3de & 14:22:13.503 & +52:17:09.71 & 2004-04-27 & 7200& N+S & 720 & 0.57 & 24.01 & 651 \\ 
04D3fj & 14:19:50.703 & +52:41:31.84 & 2004-04-28 & 7200& N+S & 720 & 0.72 & 24.19 & 108 \\ 
04D3fq & 14:16:57.906 & +52:22:46.53 & 2004-04-26 & 5400& N+S & 720 & 0.97 & 23.04 & 412 \\ 
04D3gu & 14:22:07.359 & +52:38:54.60 & 2004-05-22 & 4800& C & 720 & 1.03 & 22.60 & 6 \\ 
04D3gx & 14:20:13.678 & +52:16:58.60 & 2004-05-21 & 7200& N+S & 720 & 0.40 & 24.38 & 2439 \\ 
04D3hn & 14:22:06.878 & +52:13:43.46 & 2004-05-22 & 4800& C & 720 & 0.47 & 22.94 & 14 \\ 
04D3kr & 14:16:35.937 & +52:28:44.20 & 2004-06-16 & 2400& C & 680 & 0.82 & 21.60 & 156 \\ 
04D3lp & 14:19:50.927 & +52:30:11.85 & 2004-05-27 & 5400& N+S & 720 & 0.52 & 24.45 & 78 \\ 
04D3lu & 14:21:08.009 & +52:58:29.74 & 2004-06-23 & 3600& N+S & 720 & 0.84 & 23.49 & 17 \\ 
04D3mk & 14:19:25.830 & +53:09:49.56 & 2004-06-19 & 4320& N+S & 720 & 1.25 & 23.27 & 86 \\ 
04D3ml & 14:16:39.107 & +53:05:35.66 & 2004-06-20 & 3600& N+S & 720 & 0.41 & 23.90 & 2252 \\ 
04D3nc & 14:16:18.224 & +52:16:26.09 & 2004-07-13 & 2400& C & 720 & 0.66 & 23.39 & 149 \\ 
04D3nh & 14:22:26.729 & +52:20:00.92 & 2004-06-23 & 1800& C & 680 & 0.80 & 21.66 & 199 \\ 
04D3nq & 14:20:19.193 & +53:09:15.90 & 2004-07-14 & 1500& C & 680 & 0.61 & 21.03 & 10000 \\ 
04D3nr & 14:22:38.526 & +52:38:55.89 & 2004-07-15 & 7200& N+S & 720 & 0.69 & 24.40 & 2300 \\ 
04D3ny & 14:18:56.332 & +52:11:15.06 & 2004-07-10 & 5400& N+S & 720 & 0.79 & 23.23 & 203 \\ 
04D3oe & 14:19:39.381 & +52:33:14.21 & 2004-07-11 & 3600& N+S & 720 & 0.62 & 23.21 & 28 \\ 
04D3og & 14:20:39.748 & +53:01:15.02 & 2004-07-19 & 2700& C & 720 & 1.10 & 21.81 & 151 \\ 
04D3pd & 14:22:33.506 & +52:13:47.77 & 2004-07-18 & 3600& N+S & 720 & 0.83 & 23.34 & 72 \\ 
04D4dm & 22:15:25.470 & -17:14:42.71 & 2004-07-18 & 3600& N+S & 720 & 0.96 & 23.55 & 123 \\ 
04D4ec & 22:16:29.286 & -18:11:04.13 & 2004-07-19 & 3600& N+S & 720 & 0.70 & 23.08 & 37 \\ 
04D4ft & 22:14:31.097 & -17:40:19.74 & 2004-08-12 & 3600& C & 720 & 0.67 & 23.21 & 78 \\ 
04D4gg & 22:16:09.268 & -17:17:39.98 & 2004-08-16 & 3600& C & 720 & 0.81 & 23.03 & 30 \\ 
04D4hu & 22:15:36.193 & -17:50:19.81 & 2004-09-18 & 5400& N+S & 720 & 0.52 & 23.45 & 84 \\ 
04D4hx & 22:13:40.587 & -17:23:03.35 & 2004-09-16 & 5400& N+S & 720 & 0.73 & 24.15 & 25 \\ 
04D4ic & 22:14:21.841 & -17:56:36.43 & 2004-09-12 & 5160& N+S & 720 & 0.69 & 23.34 & 231 \\ 
04D4ih & 22:17:17.041 & -17:40:38.74 & 2004-10-07 & 5400& N+S & 720 & 0.52 & 23.95 & 41 \\ 
04D4ii & 22:15:55.645 & -17:39:27.09 & 2004-09-15 & 7200& N+S & 720 & 0.45 & 23.86 & 57 \\ 
04D4im & 22:15:00.885 & -17:23:45.84 & 2004-10-10 & 7200& N+S & 720 & 0.44 & 23.23 & 18 \\ 
04D4jy & 22:13:51.605 & -17:24:18.13 & 2004-10-14 & 8881& N+S & 720 & 0.77 & 24.08 & 1779 \\ 
04D4kn & 22:15:04.324 & -17:19:45.05 & 2004-10-19 & 5400& N+S & 720 & 0.60 & 23.74 & 97 \\ 
\enddata
\tablenotetext{a}{Exposure time in seconds.}
\tablenotetext{b}{Observing mode: nod and shuffle or classical.}
\tablenotetext{c}{Central wavelength of observing setup in nm.}
\tablenotetext{d}{Image Quality in arcseconds.}
\tablenotetext{e}{i$\arcmin$(AB) magnitude at time of spectroscopy.}
\tablenotetext{f}{Percent increase in a $1.12\arcsec$ diameter
  aperture at position of candidate at time of spectroscopy compared
  to the flux at the same position in the reference image.}
\tablenotetext{g}{Observed at Gemini-S.}
\tablecomments{SNe observed using Gemini GMOS during semesters 2003B, 2004A, and 2004B.  Observations from Gemini-N except where noted.  
Observational setup: R400 Grating, 0.75\arcsec slit, binning $2\times
2$.  One observation is not listed in this table, SNLS 04D3bf, because it was observed at Gemini-S after the SN had faded to get the host redshift.
}
\end{deluxetable}
\clearpage

\begin{deluxetable}{llllcrrrll}
\tablewidth{0pt}
\tablecaption{Derived properties for SNLS SN Candidates Observed with
  Gemini\label{der-table}.}
\tablehead{
\colhead{SN} &
\colhead{z} & 
\colhead{$\pm$} &
\colhead{Type} & 
\colhead{CI\tablenotemark{a}} & 
\colhead{$\tau_{\rm LC}$\tablenotemark{b}} & 
\colhead{$\pm$}&
\colhead{$\tau_{\rm spec}$\tablenotemark{c}} & 
\colhead{z from}
}
\startdata
03D1as & 0.872 & 0.001 & SN: & 2 & \nodata & \nodata & \nodata  & \ion{O}{2}: \\ 
03D1ax & 0.496 & 0.001 & SN Ia & 5 & -3.0 & 0.1 & -3.3  & H\&K \\ 
03D1bk & 0.8650 & 0.0005 & SN Ia & 5 & -6.3 & 0.1 & -2.1  & H\&K, H$\beta$, H$\gamma$ \\ 
03D1cj & 0.364 & 0.001 & SN Ib/c: & 1 & \nodata & \nodata & \nodata  & H$\alpha$, \ion{O}{3} \\ 
03D1cm & 0.87 & 0.02 & SN Ia & 4 & -5.0 & 0.3 & 2.4  & SN \\ 
03D1co & 0.68 & 0.01 & SN Ia & 5 & -4.8 & 0.3 & -1.2  & SN \\ 
03D1ew & 0.868 & 0.001 & SN Ia & 5 & 0.9 & 0.4 & -0.2  & \ion{O}{2} \\ 
03D1fp & 0.270 & 0.001 & SN IIb & 0 & \nodata & \nodata & \nodata  & H$\alpha$, H$\beta$, \ion{N}{2}, \ion{O}{2}, \ion{S}{2} \\ 
03D1fq & 0.80 & 0.02 & SN Ia & 4 & -1.6 & 0.3 & 2.2  & SN \\ 
03D4cj & 0.27 & 0.01 & SN Ia & 5 & -9.7 & 0.1 & -6.9  & SN \\ 
03D4ck & 0.189 & 0.001 & SN IIn & 0 & \nodata & \nodata & \nodata  & SN \\ 
03D4cn & 0.818 & 0.001 & SN Ia & 4 & -0.5 & 0.6 & 2.8  & \ion{O}{2}, \ion{O}{3}, H$\beta$ \\ 
03D4cy & 0.9271 & 0.0005 & SN Ia & 4 & 5.3 & 0.3 & 7.7  & \ion{O}{2} \\ 
03D4cz & 0.695 & 0.001 & SN Ia & 4 & 9.4 & 0.1 & 8.7  & H\&K, G-band \\ 
03D4fd & 0.791 & 0.003 & SN Ia & 5 & -1.6 & 0.3 & -1.5  & \ion{O}{2} \\ 
03D4fe & ? & \nodata & SN: & 1 & \nodata & \nodata & \nodata  & \nodata \\ 
03D4gl & 0.56 & 0.01 & SN Ia & 4 & -9.0 & 0.2 & -6.0  & SN \\ 
04D1de & 0.7677 & 0.0002 & SN Ia* & 3 & -6.8 & 0.1 & -4.5  & \ion{O}{2}, \ion{O}{3}, H$\beta$, poss  H\&K \\ 
04D1dr & 0.6414 & 0.0003 & SN: & 2 & \nodata & \nodata & \nodata  & \ion{O}{2}, \ion{O}{3}, H$\beta$, H$\gamma$ \\ 
04D1hd & 0.3685 & 0.0005 & SN Ia & 5 & -4.8 & 0.1 & -4.0  & \ion{O}{3}, weak \ion{O}{2} \\ 
04D1ho & 0.7012 & 0.0004 & SN & 2 & \nodata & \nodata & \nodata  & \ion{O}{2}, \ion{O}{3}, H\&K, H$\beta$, H$\gamma$ \\ 
04D1hy & 0.85 & 0.02 & SN Ia & 5 & -3.5 & 0.3 & -3.2  & SN \\ 
04D1jf & 0.3800 & 0.0002 & SN: & 2 & \nodata & \nodata & \nodata  & \ion{O}{2}, \ion{O}{3}, H$\beta$ \\ 
04D1ln & 0.2072 & 0.0002 & SN II-P & 0 & \nodata & \nodata & \nodata  & H$\alpha$, \ion{N}{2}, \ion{S}{2} \\ 
04D1ow & 0.93 & 0.02 & SN Ia & 4 & 2.6 & 0.2 & -0.8  & SN \\ 
04D2aa\tablenotemark{g} & ? & \nodata & SN: & 2 & \nodata & \nodata & \nodata  & \nodata \\ 
04D2ad\tablenotemark{g} & 0.6802 & 0.0002 & SN: & 2 & \nodata & \nodata & \nodata  & \ion{O}{2}, \ion{O}{3}, H$\beta$ \\ 
04D2ae\tablenotemark{g} & 0.843 & 0.001 & SN Ia & 4 & 0.0 & 0.1 & -1.5  & H\&K \\ 
04D3aa & 0.2045 & 0.0002 & SN II & 0 & \nodata & \nodata & \nodata  & H$\alpha$, H$\beta$, \ion{O}{3}, \ion{S}{2} \\ 
04D3ae & 0.217 & 0.001 & SN II & 0 & \nodata & \nodata & \nodata  & H$\alpha$, \ion{O}{3} \\ 
04D3ax & 0.3558 & 0.0002 & SN II: & 1 & \nodata & \nodata & \nodata  & H$\alpha$, H$\beta$, \ion{O}{3} \\ 
04D3bf & 0.1560 & 0.0005 & SN Ia & 5 & \nodata & \nodata & 15.1  & \ion{O}{2}, \ion{O}{3}, \ion{S}{2}, H$\alpha$, H$\beta$ \\ 
04D3dd & 1.01 & 0.02 & SN Ia & 4 & 2.9 & 0.3 & -1.4  & SN \\ 
04D3de & ? & \nodata & SN II-P & 0 & \nodata & \nodata & \nodata  & \nodata \\ 
04D3fj & ? & \nodata & SN: & 1 & \nodata & \nodata & \nodata  & \nodata \\ 
04D3fq & 0.73 & 0.01 & SN Ia & 4 & 0.8 & 0.3 & 3.6  & SN \\ 
04D3gu & 0.748 & 0.001 & SN: & 2 & \nodata & \nodata & \nodata  & H\&K, Balmer, weak  \ion{O}{2} \\ 
04D3gx & 0.91 & 0.02 & SN Ia* & 3 & 10.8 & 0.2 & 8.3  & SN \\ 
04D3hn & 0.5516 & 0.0003 & SN Ia & 5 & 7.2 & 0.1 & 4.9  & H\&K, Balmer \\ 
04D3kr & 0.3373 & 0.0002 & SN Ia & 5 & 4.4 & 0.1 & 1.7  & \ion{O}{2}, \ion{O}{3}, H$\beta$ \\ 
04D3lp & 0.983 & 0.001 & SN Ia* & 3 & 1.0 & 0.2 & -1.2  & \ion{O}{2} \\ 
04D3lu & 0.8218 & 0.0002 & SN Ia & 4 & 5.5 & 0.1 & 5.6  & H\&K \\ 
04D3mk & 0.813 & 0.001 & SN Ia & 5 & -2.4 & 0.1 & 1.2  & \ion{O}{2}, H\&K \\ 
04D3ml & 0.95 & 0.02 & SN Ia & 4 & -1.8 & 0.3 & 1.1  & SN \\ 
04D3nc & 0.817 & 0.001 & SN Ia* & 3 & 7.2 & 0.2 & \nodata  & poss  \ion{O}{2} \\ 
04D3nh & 0.3402 & 0.0002 & SN Ia & 5 & 3.5 & 0.1 & 3.7  & H$\alpha$, H$\beta$, \ion{O}{2}, poss H\&K \\ 
04D3nq & 0.22 & 0.01 & SN Ia & 5 & 8.1 & 0.1 & 8.8  & SN \\ 
04D3nr & 0.96 & 0.02 & SN Ia* & 3 & 8.1 & 0.3 & 3.9  & SN \\ 
04D3ny & 0.81 & 0.02 & SN Ia & 5 & 1.6 & 0.2 & 2.5  & SN \\ 
04D3oe & 0.756 & 0.001 & SN Ia & 4 & 1.4 & 0.1 & 1.8  & H\&K \\ 
04D3og & 0.352 & 0.001 & SN & 2 & \nodata & \nodata & \nodata  & H$\alpha$, H$\beta$, \ion{O}{2}, \ion{N}{2}, \ion{S}{2} \\ 
04D3pd & 0.760 & 0.001 & SN: & 2 & \nodata & \nodata & \nodata  & H$\beta$, \ion{O}{2}, \ion{O}{3} \\ 
04D4dm & 0.811 & 0.001 & SN Ia & 4 & 2.7 & 0.2 & -0.7  & \ion{O}{2}, \ion{O}{3} \\ 
04D4ec & 0.593 & 0.001 & SN: & 2 & \nodata & \nodata & \nodata  & \ion{O}{2}, \ion{O}{3}, H$\beta$, H$\gamma$ \\ 
04D4ft & 0.2666 & 0.0002 & SN & 2 & \nodata & \nodata & \nodata  & \ion{O}{3}, H$\alpha$, H$\beta$ \\ 
04D4gg & 0.4238 & 0.0004 & SN Ia & 5 & -9.7 & 0.1 & -10.0  & \ion{O}{2}, \ion{O}{3}, H$\alpha$, H$\beta$, H$\gamma$ \\ 
04D4hu & 0.7027 & 0.0003 & SN Ia & 5 & 5.2 & 0.5 & 5.0  & \ion{O}{2}, \ion{O}{3}, H$\beta$ \\ 
04D4hx & 0.545 & 0.005 & SN: & 2 & \nodata & \nodata & \nodata  & 4000\AA\ break, poss  H\&K \\ 
04D4ic & 0.68 & 0.02 & SN Ia & 4 & 2.5 & 0.3 & 4.8  & SN \\ 
04D4ih & 0.934 & 0.001 & SN Ia* & 3 & 8.7 & 0.2 & 5.6  & \ion{O}{2}, H\&K, some Balmer \\ 
04D4ii & 0.866 & 0.001 & SN Ia & 5 & -4.6 & 0.2 & -3.5  & \ion{O}{2} \\ 
04D4im & 0.7510 & 0.0005 & SN Ia & 4 & 0.7 & 0.2 & 0.2  & H\&K, H$\delta$ \\ 
04D4jy & 0.93 & 0.02 & SN Ia* & 3 & 2.0 & 0.4 & 0.0  & SN \\ 
04D4kn & 0.9095 & 0.0005 & SN: & 2 & \nodata & \nodata & \nodata  & \ion{O}{2}, H\&K, Balmer \\ 
\enddata
\tablenotetext{a}{SN Ia confidence index: (5) Certain Ia; (4) Highly probable
  Ia; (3) Probable Ia; (2) Unidentified; (1) Probably not a Ia; (0) Certainly not a Ia.}
\tablenotetext{b} {Epoch determined from lightcurve:  Rest frame epoch
  at which spectroscopy was taken relative to rest frame B-band maximum light.}
\tablenotetext{c}{Epoch determined from spectroscopic fit.
The uncertainty on this value is 2.5 days.}
\tablecomments{Derived properties for SNe listed in
  Table~\ref{obs-table}.  A colon denotes uncertainty.}
\end{deluxetable}
\clearpage

\begin{deluxetable}{cccc}
\tablecaption{Ia Confidence Index and Host Type\label{class-host-table}.}
\tablehead{
\colhead{CI\tablenotemark{a}} & 
\colhead{Abs.\tablenotemark{b}} &
\colhead{Emis.\tablenotemark{c}} & 
\colhead{None\tablenotemark{d}} 
}
\startdata
5 & 2 & 11 & 5\\
4 & 5 & 3  & 8\\
3 & 0 & 4  & 3\\
2 & 1 & 11  & 1\\
1 & 0 & 2  & 2\\
0 & 0 & 4  & 2\\
\enddata
\tablenotetext{a}{Ia confidence index: (5) Certain Ia; (4) Highly probable Ia; (3) Probable Ia; (2) Unidentified; (1) Probably not a Ia; (0) Certainly not a Ia.}
\tablenotetext{b}{Hosts have only absorption lines.}
\tablenotetext{c}{Hosts have emission features.}
\tablenotetext{d}{Candidates with no host features, either because
  there was no apparent host or because the slit could not be placed
  through the galaxy.}
\tablecomments{Note that probable and certain core
  collapse SNe (index 0 and 1) do not occur in absorption
  line galaxies.}
\end{deluxetable}

\end{document}